\documentclass[%
reprint,
amsmath,amssymb,
aps,
prl,
]{revtex4-1}

\usepackage[utf8]{inputenc}
\usepackage{graphicx}% Include figure files
\usepackage{dcolumn}% Align table columns on decimal point
\usepackage{bm}% bold math
\usepackage{tikz-feynman}
\usepackage{subcaption}
\usepackage[colorlinks=true,linktocpage=true,linkcolor=blue,citecolor=blue,allcolors=blue]{hyperref}% add hypertext capabilities
\usepackage{empheq}
\usepackage{multirow}
\usepackage{booktabs}

\usepackage{color}
\usepackage{xspace}
\usepackage[capitalize]{cleveref}
\usepackage{bm}
\newcommand{\REM}[1]{}

\definecolor{magenta}{cmyk}{0,1,0,0}

\begin{document}
	
	%\preprint{APS/123-QED}
	
\title{Formation, dissociation and regeneration of charmonia \\
within microscopic Langevin simulations}

\author{Naomi Oei}
\thanks{Corresponding author}
\email{oei@itp.uni-frankfurt.de}
\affiliation{Institut f\"ur Theoretische Physik, Johann Wolfgang Goethe-Universit\"at, Max-von-Laue-Strasse 1, 60438 Frankfurt am Main, Germany}% 

 %\author{Nadja Krenz}
	%\affiliation{Institut f\"ur Theoretische Physik, Johann Wolfgang Goethe-Universit\"at, Max-von-Laue-Strasse 1, 60438 Frankfurt am Main, Germany}%

\author{Juan Torres-Rincon}
\affiliation{Departament de F\'isica, Qu\`antica i Astrof\'isica and Institut de Ci\`encies del Cosmos (ICCUB), Facultat de F\'isica, Universitat de Barcelona, 08028 Barcelona, Spain}%

\author{Nadja Krenz}
\thanks{left academia}
\affiliation{Institut f\"ur Theoretische Physik, Johann Wolfgang Goethe-Universit\"at, Max-von-Laue-Strasse 1, 60438 Frankfurt am Main, Germany}%
	
\author{Hendrik van Hees}
%\email{hees@itp.uni-frankfurt.de}
\affiliation{Institut f\"ur Theoretische Physik, Johann Wolfgang Goethe-Universit\"at, Max-von-Laue-Strasse 1, 60438 Frankfurt am Main, Germany}%

\author{Carsten Greiner}
%\email{carsten.greiner@itp.uni-frankfurt.de}
\affiliation{Institut f\"ur Theoretische Physik, Johann Wolfgang Goethe-Universit\"at, Max-von-Laue-Strasse 1, 60438 Frankfurt am Main, Germany}%

\date{\today}
	
\begin{abstract}

We present a microscopic dynamical model to study the formation, dissociation, and recombination processes of charmonium states in a heat bath at constant temperature and volume. Within this classical approach, heavy quarks are described as Brownian particles in a background medium of light constituents and can therefore be modeled by a Fokker-Planck equation with constant transport coefficients, which is then implemented through relativistic Langevin simulations. The heavy quarks interact classically via a Coulomb-like screened potential to form a bound state if the relative energy of the pair becomes negative. Dissociation of bound states is possible as a result of screening effects on the potential as well as through scatterings with plasma particles. We demonstrate the full equilibration of the system and show that the resulting equilibrium charmonium yields are in accordance with the Statistical Hadronization Model.
\end{abstract}
	
%\keywords{Suggested keywords}%Use showkeys class option if keyword
%display desired
\maketitle
	
%%%%%%%%%%%%%%%%%%%%%%%%%%%%%%%%%%%%%%%%%%%%%%%%%%%%%%%%%%%%%%%%%%%%%%%%%%%%%%%
 
\section{Introduction}

The investigation of heavy-quark and quarkonium states have been considered an important probe to study properties of the medium created in relativistic heavy-ion collisions (RHICs) since the prediction of $J/\psi$ suppression by Matsui and Satz in 1986~\cite{MatsuiSatz}. They showed that under sufficiently high densities or temperatures in order to create a quark-gluon plasma (QGP), the interaction to form a bound state of a charm and an anticharm quark is suppressed due to the screening of the color charges of the unconfined quarks and gluons in the medium. This effect was experimentally observed and measured, first at SPS~\cite{NA50:1997hlx,NA50:2000brc,NA50:2004sgj}
and additionally later at RHIC~\cite{PHENIX:2006gsi,PHENIX:2011img,STAR:2009irl}.

Pairs of heavy quarks and antiquarks are produced in primordial hard scatterings and their number is conserved throughout the whole medium evolution. Therefore information about their interaction history with the QGP can be obtained from measuring hadrons containing heavy quarks. Additionally, dissociation of quarkonium states may occur through quasi-elastic scatterings with medium particles and, at sufficiently high collision energies, these processes are possibly countered by regeneration of heavy quarks and antiquarks during the medium evolution~\cite{Cao:2013ita,Cao:2015hia,Cao:2016gvr,Cao:2017hhk,Gossiaux:2009mk,Nahrgang:2013saa,He:2011qa,vanHees:2007me,Du:2022uvj,Plumari:2012ep,Scardina:2017ipo,Cassing:2008sv,Song:2015sfa,Pooja:2024rnn} (for a summary of models and their comparison see Refs.~\cite{Cao:2018ews, Andronic:2024oxz}). A successful way to predict quarkonium yields measured in experiments at the hadronization phase boundary is the Statistical Hadronization Model (SHM)~\cite{Andronic:2003zv,Andronic:2017pug,Andronic:2021erx}. This description neglects the dynamics that might occur during the medium evolution after the initial collision and instead focuses on describing the production of hadrons at the freeze-out stage of the QGP in the framework of statistical mechanics. This process is viewed as an instantaneous phenomenon governed by the chemical freeze-out temperature and the baryochemical potential, under the assumption of a statistically equilibrated source volume which can be described by a grand-canonical ensemble. 

In this work we propose a classical description of the heavy-quark motion, starting from a Fokker-Planck equation which is derived from the Boltzmann equation of the heavy-quark phase-space distribution and implemented through relativistic Langevin simulations. Within this approach the heavy quarks are approximated as Brownian particles and their motion results from random kicks due to quasi-elastic scatterings with light medium particles. The heavy quarks interact over a complex potential of which the real screened Coulomb-like part describes an attractive force between a heavy-quark-antiquark pair in order to form bound states, while the imaginary part takes interactions with the medium particles into account. In that way, the formation of charmonia as well as dissociation and recombination processes of bound states in the QGP can be described.
 
The paper is organized as follows: In Sec.~\ref{sec:FP_Langevin} the formalism of the Fokker-Planck equation is introduced and the realization of the heavy-quark motion in the form of a Langevin equation is described. Afterwards, Sec.~\ref{sec:HQpotential} explains the potential between heavy quarks and antiquarks, which is able to form bound states. This (complex) potential can describe dissociation processes of bound states via its imaginary part, from which the drag coefficient is extracted. Section~\ref{sec:BSformation} focuses on the formation of bound states, the binding energy of a charmonium state as well as the time evolution of the bound states under different conditions. Furthermore, the relaxation time of the charmonium states in connection with the drag coefficient is studied in Sec.~\ref{sec:eq_time} Finally, in Sec.~\ref{sec:sectionSHM} the charmonium yields resulting from our model are compared to the predictions of the SHM. Finally, we present our conclusions and outlook in Sec.~\ref{sec:conclusions}.

\section{Fokker-Planck equation and Langevin simulations} \label{sec:FP_Langevin}

The dynamics of heavy quarks in a thermal bath can be described by Brownian motion due to their large mass relative to the masses and the typical momentum exchanges with the light constituents of the QGP. The motion of heavy particles is viewed as random kicks resulting from momentum exchanges from quasi-elastic scatterings with light particles of the background medium. Under the assumption of local thermal equilibrium such an approach can be realized with a Fokker-Planck equation, describing the time evolution of the probability density of the heavy quarks under the influence of a drag and random forces.
 %model the propagation of the heavy quarks through the medium.
This equation can be derived from the relativistic Boltzmann equation for heavy-quark phase space distribution,
\begin{equation}
   \left [ \frac{\partial}{\partial t} + \frac{\bm{p}}{E} \cdot \frac{\partial}{\partial \bm{x}} + \bm{F}    \cdot \frac{\partial}{\partial \bm{p}} \right ] f_Q(t, \bm{p}, \bm{x}) = C[f_Q] \ ,
\label{Boltzmann}
\end{equation}
where the left-hand side contains the advective term and the forces acting on the heavy quark, while the collision integral on the right-hand side represents the local interactions of heavy quarks with the light particles of the medium,
\begin{equation}
    C[f_Q] = \int d^3\bm{k} \ [ \omega(\bm{p} + \bm{k}, \bm{k})f_Q(\bm{p}+ \bm{k}) - \omega(\bm{p}, \bm{k})f_Q(\bm{p}) ] \ , 
    \label{col_int}
\end{equation}
with the transition rate $\omega(\bm{p}, \bm{k})$ of a scattering process where the heavy-quark momentum changes from $\bm{p}$ to $\bm{k}$, by either gaining or losing additional momentum through interactions with the medium particles. 

In the limit of small exchanged momentum $\bm{k}$ as compared to $\bm{p}$, the collision integral can be expanded reducing the Boltzmann equation to a Fokker-Planck equation~\cite{svetitsky, pitaevskii2012physical, Langevin-vanHees-Rapp, rapp-vanHees}. Assuming, in addition, homogeneity and the absence of mean-field effects one gets,
\begin{equation}
    \frac{\partial}{\partial t} f_Q(\bm{p},t) = \frac{\partial}{\partial p_i} \Bigg\{ A_i(\bm{p})f_Q(\bm{p},t) + \frac{\partial}{\partial p_j} [B_{ij}(\bm{p})f_Q(\bm{p},t)] \Bigg\}     \ ,
    \label{fokkerplanck}
\end{equation}
where $i$ and $j$ are spatial indices and the transport coefficients $A_i(\bm{p})$ and $B_{ij}(\bm{p})$ follow from the evaluation of the first two moments of the collision integral,
\begin{align}
    A_i(\bm{p}) & = \int d^3\bm{k} \ \omega(\bm{p}, \bm{k})k_i \ , \\
    B_{ij}(\bm{p}) & = \frac{1}{2} \int d^3 \bm{k} \ \omega(\bm{p}, \bm{k}) k_i k_j \ ,
\end{align}
called the drag force and diffusion coefficients, which encode the forces acting on the heavy quark and its interactions with the medium. Imposing that the solution of the Fokker-Planck equation converges to the relativistic Boltzmann-J\"uttner distribution in equilibrium, the two coefficients are connected via the dissipation-fluctuation theorem~\cite{Langevin-vanHees-Rapp},
\begin{equation}
    A_i(\bm{p},T) = B_{ij}(\bm{p},T) \frac{1}{T} \frac{\partial E(p)}{\partial p_j} - \frac{\partial B_{ij}(\bm{p},T)}{\partial p_j}\ ,
    \label{diss-fluct}
\end{equation}
where $E(p)=\sqrt{p^2+M_c^2}$ is the relativistic energy of the charm quark with mass $M_c$.This was first derived in \cite{pitaevskii2012physical} for the non-relativistic case.

In an isotropic medium the drag force and diffusion coefficients reduce to three scalar quantities, $A(\bm{p},T), B_0(\bm{p},T)$ and $B_1(\bm{p},T)$,
\begin{align}
    A_i(\bm{p},T) & = A(\bm{p},T) p_i \ , \label{eq:Ai} \\
    B_{ij}(\bm{p},T) & = B_0 (\bm{p},T) \left( \delta_{ij} - \frac{p_i p_j}{p^2} \right)  + B_1(\bm{p},T) \frac{p_i p_j}{p^2} \ \label{eq:Bij} .
\end{align}

Following Ref.~\cite{svetitsky} we consider the static limit $\bm{p} \xrightarrow[]{} 0$, where the Fokker-Planck equation reduces to the Rayleigh's equation with a diagonal approximation of the diffusion coefficient
% Since the charm quark mass exceeds the temperature of the system, both are chosen to be constant in this description,  
\begin{equation}
 B_0(\bm{p},T) = B_1(\bm{p},T) \equiv D( \bm{p},T) \ . \label{eq:static}
\end{equation}

Assuming these conditions the fluctuation-dissipation relation in~\eqref{diss-fluct} reduces to~\cite{Langevin-vanHees-Rapp},
\begin{align}
   A(p) & = \frac{1}{E(p)} \left( \frac{1}{T} - \frac{\partial}{\partial E} \right) D[E(p)] \ , \label{eq:flucdissimp}
\end{align}    
where the momentum dependence of the diffusion coefficient enters through the particle energy. To ease the notation, in the following we will omit the $T$ argument in the transport coefficients.

In our following numerical approach, we consider a microscopical description of the system. This is realized through the relativistic Langevin equation \cite{PhysRevC.73.034913} \cite{DUNKEL20091}, and serves as the starting point for the practical simulation of heavy-quark propagation. The stochastic equations of motion for the heavy quarks are defined by the Langevin update steps, describing the change of position and momentum discretized in a time interval due to random momentum kicks from quasi-elastic scatterings with medium particles,
% \begin{empheq}[left=\empheqlbrace]{align}
    %    \frac{d\bm{x}}{d t} &= \frac{\bm{p}}{E(p)}  \ , \\
    % \frac{d\bm{p}}{d t} &= - \gamma \bm{p} + \bm{\xi} \ ,
% \end{empheq}
% \begin{align}
%     \frac{dp^\mu}{d \tau} &= - \gamma p^\mu + \xi^\mu \ ,
% \end{align}
% with the friction coefficient $\gamma$ and the stochastic force $\bm{\xi}$, where the probability for the directions of a random kick is isotropic and with zero average,
% \begin{equation}
%     \langle \xi_j(t)\rangle = 0\ .
% \end{equation}
\begin{empheq}[left=\empheqlbrace]{align}
    dx_j & =  \frac{p_j}{E(p)} dt \  ,   \label{xupdate} \\
    dp_j & =  -\gamma(p) p_j dt + \sqrt{dt} C_{jk}(p) \rho_k \ , \label{momupdate}
\end{empheq}
where repeated spatial indices must be summed over. $\gamma(p)$ is the deterministic friction coefficient, and $C_{jk}(p)$ parametrizes the strength of the stochastic force. Such a noise is represented by the random variable $\bm{\rho}$ following a Gaussian normal distribution, 
\begin{align}
   P(\bm{\rho}) = (2\pi)^{-\frac{3}{2}} e^{{-\frac{\bm{\rho}^2}{2}}}   \ ,
\end{align}
where its moments satisfy, 
\begin{equation}
    \langle \rho_k \rangle  = 0 \ , \quad 
    \langle \rho_j \rho_k \rangle  = \delta_{jk} \ .
\end{equation}
describing what is called a white noise.

The numerical implementation of the random term in Eq.~\eqref{momupdate} is not unambiguous, and one needs to specify at which momentum step the 
covariance of the stochastic force $C_{jk}$ is applied. This is parametrized by a scalar factor $\zeta$, 
%the stochastic process is still dependent on the specific choice of the momentum argument of the covariance matrix Cjk, with the possibilites of the pre-point, mid-point and post-point interpretation of the stochastic integral. The various interpretations can be summarized by determining the momentum argument in the covariance matrix,
\begin{equation}
    C_{jk} \xrightarrow{} C_{jk}(\bm{p}+ \zeta d\bm{p}) \ ,
\end{equation}
with the typical choices $\zeta = \{ 0, \frac{1}{2}, 1 \}$ for pre-, mid- and post-point realizations, respectively. %In this work the post-point scheme will be used, i.e.
%\begin{equation}
%    C_{jk} \xrightarrow{} C_{jk}(t, \bm{x}, \bm{p}+ d\bm{p})\ .
%\end{equation} 
% which reduces the general equilibrium condition Eq.~(\ref{diss-fluct}) to
% \begin{equation}
%     D[E(p)] = \gamma E(p) T \ ,
% \end{equation}
% where the diffusion coefficient $D$ receives a momentum dependence.

The phase-space distribution which follows from the previous Langevin equation satisfies a Fokker-Planck equation, which can be derived by computing the time evolution of the average of the phase-space function. The result from Ref.~\cite{rapp-vanHees} is,
\begin{align}
\frac{\partial f(t, \bm{p})}{\partial t} & = \frac{\partial}{\partial p_j} \left[ \left( \gamma(p) p_j - \zeta C_{lk}(\bm{p}) \frac{\partial C_{jk}(\bm{p})}{\partial p_l} \right) f(t, \bm{p}) \right] \nonumber \\
& + \frac{1}{2} \frac{\partial^2}{\partial p_j \partial p_k} \left[ C_{jl}(\bm{p}) C_{kl}(\bm{p}) f(t, \bm{p}) \right] \ .  \label{eq:FKderived}
\end{align}

At this point we can do the matching between Eq.~\eqref{eq:FKderived} and the form given in Eq~\eqref{fokkerplanck}, that was obtained from the Boltzmann equation. The relations between the two sets of coefficients are~\cite{rapp-vanHees}, 
\begin{align}
A(p) \ p_j &= \gamma (p) \ p_j - \zeta C_{lk} (\bm p) \frac{\partial C_{jk} (\bm p)}{\partial p_l} \ , 
\label{Apj}\\
B_{jk}(\bm p) &= \frac{1}{2} C_{jl}(\bm p) C_{kl}(\bm p) \ . \label{Bpj}
\end{align}

Equation~\eqref{Bpj} can be rewritten by applying the longitudinal/transverse projections made in Eq.~\eqref{eq:Bij} for an isotropic medium. One gets,
\begin{align}
    C_{jk} (\bm p) & = \sqrt{2B_0(p)} \left( \delta_{jk} - \frac{p_j p_k}{p^2} \right) + \sqrt{2B_1 (p)} \ \frac{p_j p_k}{p^2} \ . \label{eq:CB}
\end{align}
Taking, in addition, the static limit (for which $B_0(p)=B_1(p) \rightarrow D(p)$) we arrive at the simple relation
\begin{equation}
    C_{jk} (p) = \sqrt{2D [E(p)]} \ \delta_{jk} \ . \label{eq:CD}
\end{equation}

Inserting (\ref{eq:CD}) into Eq.~\eqref{Apj}, the equation that related the drag and friction forces is simplified to
\begin{equation}
A(p)= \gamma(p) - \zeta \frac{1}{E(p)} \frac{\partial D [E(p)]}{\partial E} \ ,\label{eq:AGD}
\end{equation}
where we observe that for a general $\zeta$, the relation depends also on the diffusion coefficient $D(p)$.

Finally, we will find a simple relation between the friction from in the Langevin equation $\gamma(p)$ and the diffusion coefficient $D(p)$ by equating Eq.~\eqref{eq:AGD} to the fluctuation-dissipation relation~(\ref{eq:flucdissimp}). We obtain~\cite{Langevin-vanHees-Rapp}
\begin{align}
    \gamma(p) = \frac{1}{E(p)} \left( \frac{D[E(p)]}{T} - (1-\zeta) \frac{\partial D[E(p)]}{\partial E} \right) \ ,
\end{align}
where it is a fluctuation-dissipation relation between the friction force and the diffusion coefficient (in the static limit).

In this work we will use the post-point scheme ($\zeta = 1$) for which we can reduce Eq.~(\ref{diss-fluct}) to the Einstein's relation,
\begin{equation}
    D[E(p)] = \gamma E(p) T \ . \label{eq:einstein}
\end{equation}
We will assume a momentum-independent value of $\gamma$, but a diffusion coefficient $D(p)$ that carries a momentum dependence.

Using~\eqref{eq:CD} and~\eqref{eq:einstein} we can finally write the specific form of the Langevin update steps for the heavy quarks in this description, 
% leads to the Langevin update steps which are the equations of motion for the heavy quarks, describing the change of position and momentum in a time interval due to random momentum kicks from \textcolor{blue}{quasi-elastic} scatterings with medium particles:
\begin{empheq}[left=\empheqlbrace]{align}
    dx_j & = \frac{p_j}{E(p)} \, dt \label{eq:langevin-x} \\
    dp_j & = -\gamma p_j \, dt + \sqrt{2\gamma E(p) T \, dt} \, \rho_j \ . 
    \label{eq:langevin-mom}
\end{empheq}

% While the coordinate update is straightforward, the momentum update requires a two-step computation due to the choice of the post-point scheme for the covariance matrix. Primarily, the momentum update in the pre-point scheme with $\zeta=0$ is computed, which can be utilized to calculate the momentum increment $dp_j^{\mathrm{drag}}$ with $\gamma$ and $D(p)$. Subsequently, the diffusion coefficient at $|\bm{p} + d\bm{p}|$ is evaluated in order to calculate $dp_j^{\mathrm{diffusion}} = \sqrt{2dt D(|\bm{p} + d\bm{p}|)} \ \rho_j$. Finally, the sum of $dp_j^{\mathrm{diffusion}}$ and  $dp_j^{\mathrm{drag}}$ yields the total momentum increment for one full time step. 
While the coordinate update is straightforward, the momentum update requires a two-step computation due to the choice of the post-point scheme ($\zeta=1$) for the covariance matrix. In the first step, a preliminary momentum increment, $dp_j^{\mathrm{drag}}$, is computed using $\gamma$ and $D(p)$. Subsequently, the diffusion coefficient is recalculated at the updated momentum, $|\bm{p} + d\bm{p}|$, in order to evaluate $dp_j^{\mathrm{diffusion}} = \sqrt{2dt D(|\bm{p} + d\bm{p}|)} \ \rho_j$. Finally, the total momentum increment for one full time step is obtained as the sum of $dp_j^{\mathrm{diffusion}}$ and $dp_j^{\mathrm{drag}}$. \cite{Langevin-vanHees-Rapp, Klimontovich1994, DUNKEL20091}\\

In the end Eq.~(\ref{eq:langevin-mom}) represent the well-known relativistic Langevin equation,
\begin{align}
    \frac{dp^\mu}{d t} = - \gamma p^\mu + \xi^\mu
    \label{eq:langevin}
\end{align}
with the friction coefficient $\gamma$ and the stochastic force $\xi^\mu$, where the probability for the directions of a random kick is isotropic and with zero average, 
\begin{align}
    \langle \xi_j(t) \rangle = 0 \ .
\end{align}
and for the noise correlation

\begin{align}
    \langle \xi(t_1) \xi(t_2) \rangle = 2 \gamma E T \delta(t_2 - t_1) \ .
\end{align}

In principle, the white noise could be replaced by a colored, non-Markovian noise with a non-local correlation in time~\cite{PhysRevE.107.064131,PhysRevD.108.054026}. However, such standard Langevin equations are much more numerically involved 
and thus for our present setting we will stay with the Markovian approximation, Eq.~(\ref{eq:langevin}).

\section{Heavy Quark Potential}\label{sec:HQpotential}

The potential used to describe the interaction between a heavy-quark pair allowing to form a bound state is adopted from the formalism by Blaizot {\it et al.}~\cite{Blaizot}. In this approach the dynamics of the heavy quarks are described as an Abelian Plasma, where the mass difference between heavy and light quarks is used to formulate an effective theory in which it is possible to treat the heavy quarks as non-relativistic particles in a plasma of relativistic particles, considering the plasma properties in the correlation functions. 
%Of course the restriction to abelian interactions for the description of the medium  
To derive the effective potential for the heavy quark, Ref.~\cite{Blaizot} makes use of the influence functional method, and integrate out the medium degrees of freedom to derive an effective complex potential between heavy particles,
\begin{equation}
    \mathcal{V}(r) = - \frac{g^2}{4\pi} m_D - \frac{g^2}{4\pi}  \frac{\exp(-m_D r)}{r} - i \frac{g^2T}{4 \pi}\phi(m_D r) \ , \label{eq:Blaizot}
\end{equation}
where $r$ is the relative distance between the heavy quark and antiquark. In Eq.~(\ref{eq:Blaizot}) the first term defines a self-energy contribution, which amounts in a constant potential. The real Coulomb-like second term contains the attractive potential developing between a heavy-quark pair, screened by the Debye mass $m_D$ of the light component of the gas, while the imaginary part in the third complex term encodes the momentum loss in the scatterings with medium particles. The function $\phi(m_Dr)$ is given by
\begin{equation}
        \phi(x)= 2 \int_0^\infty dz \frac{z}{(z^2+1)^2} \left[ 1- \frac{\sin(zx)}{zx} \right]\ .
\end{equation}
Within the potential in Eq.~(\ref{eq:Blaizot}) dissociation processes are possible resulting on one hand from the screening of the potential, and on the other hand from collisions with the plasma particles. To regularize the singularity of the potential at small distances, we apply a cut-off $\Lambda = 4 \ \text{GeV}$ in the momentum integral of the real part of the potential, as proposed in Ref.~\cite{Blaizot}. The strong coupling is parametrized by
\begin{equation}
    g^2 (T) = 4 \pi \alpha_s(T) = \frac{4\pi \alpha_s(T_c)}{1+ C \ln \left( T/T_c \right)} \ , 
\end{equation}
with $C=0.76$ and $T_c = 160\ \text{MeV}$. 

Typically, a value of $\alpha_s(T_c) \simeq 0.5$ is chosen for the strong coupling~\cite{Blaizot}. However, in this work we use $\alpha_s = 0.7$ as will be argued in Sec.~\ref{sec:sectionSHM}. %The value for the strong coupling $\alpha_s(T_c)$ is chosen in order to achieve an agreement in the charmonium yield with the Statistical Hadronization Model, which is explained in more detail in section VI. 
The mass of the charm quark is chosen to be $M_c=1.8\ \text{GeV}$, because this will eventually lead to a $J/\Psi$ physical mass of $3.1\ \text{GeV}$, when a heavy-quark pair is bound. 

\begin{figure}[h]
    \centering
    \includegraphics[width = 8cm]{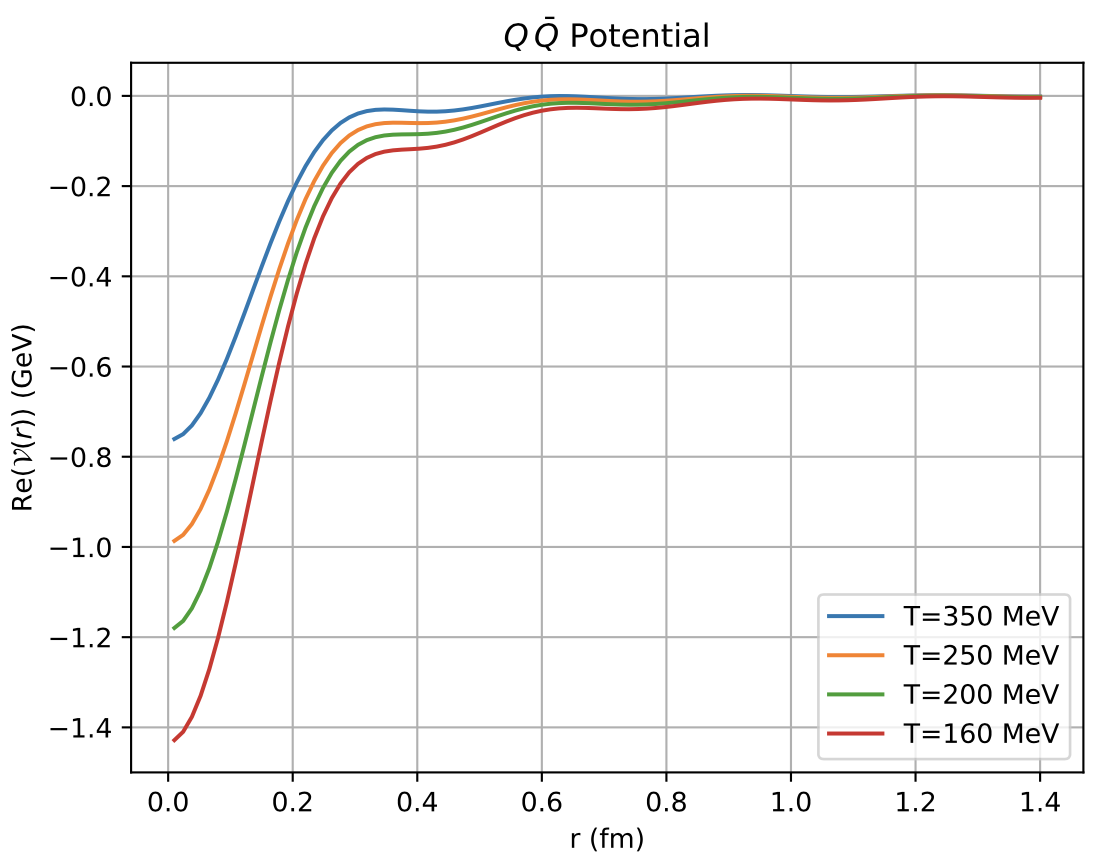}
    \caption{Real part of the complex potential from Ref.~\cite{Blaizot} as a function of the relative distance of the heavy-quark pair for different temperatures. Notice that the potential is qualitatively similar to the one in Ref.~\cite{Blaizot}, but since modify the strong-coupling $\alpha_s$ (as explained in Sec.~\ref{sec:sectionSHM}), it leads to a deeper potential for all considered temperatures.}
    \label{fig:potential}
\end{figure}

In Fig.~\ref{fig:potential} the real part of the potential (\ref{eq:Blaizot}) is displayed as a function of the relative distance of the heavy-quark pair for four different temperatures. Upon comparison of the curves at different temperatures, the screening effects of the potential become apparent, leading to a decreasing depth of the potential for higher temperatures, where the screening of the unconfined color charges in the medium hinders the formation of bound states. Therefore, the interaction leading to the formation of bound states is enhanced at lower temperatures and we will expect a higher fraction of bound states.

The drag coefficient $\gamma$ that appears in the Langevin equation of the charm quarks describes how the average momentum equilibrates through scatterings and can therefore also be interpreted as a thermal relaxation rate. It is possible to define a relaxation time $\tau_R=\gamma^{-1}$ which is the characteristic time for a heavy particle to acquire a thermal momentum. In Ref.~\cite{Blaizot}, 
$\gamma$ follows from the derivative of the imaginary part of the potential (\ref{eq:Blaizot}), and it is, in principle, a function of the relative distance of the pair. However, in our simulation this dependence will be neglected and we use a constant drag coefficient instead~\cite{Blaizot}
\begin{align}
    \gamma & = \frac{g^2}{M_c T}  \delta^{ij}  \frac{\partial^2}{\partial r_i \partial r_j} \left[ - \int^\Lambda \frac{d\bm{k}}{(2\pi)^3} e^{i \bm{k} \cdot \bm{r}}  \frac{\pi m_D^2 T}{k (k^2+m_D^2)^2} \right]_{r=0} \nonumber \\
    & = \frac{m_D^2g^2}{24\pi M_c} \left[ \ln \left( 1+ \frac{\Lambda^2}{m_D^2} \right) - \frac{\frac{\Lambda^2}{m_D^2}}{\frac{\Lambda^2}{m_D^2}+1} \right] \ .
\end{align}

\begin{figure}[h]
\centering
\includegraphics[width=8cm]{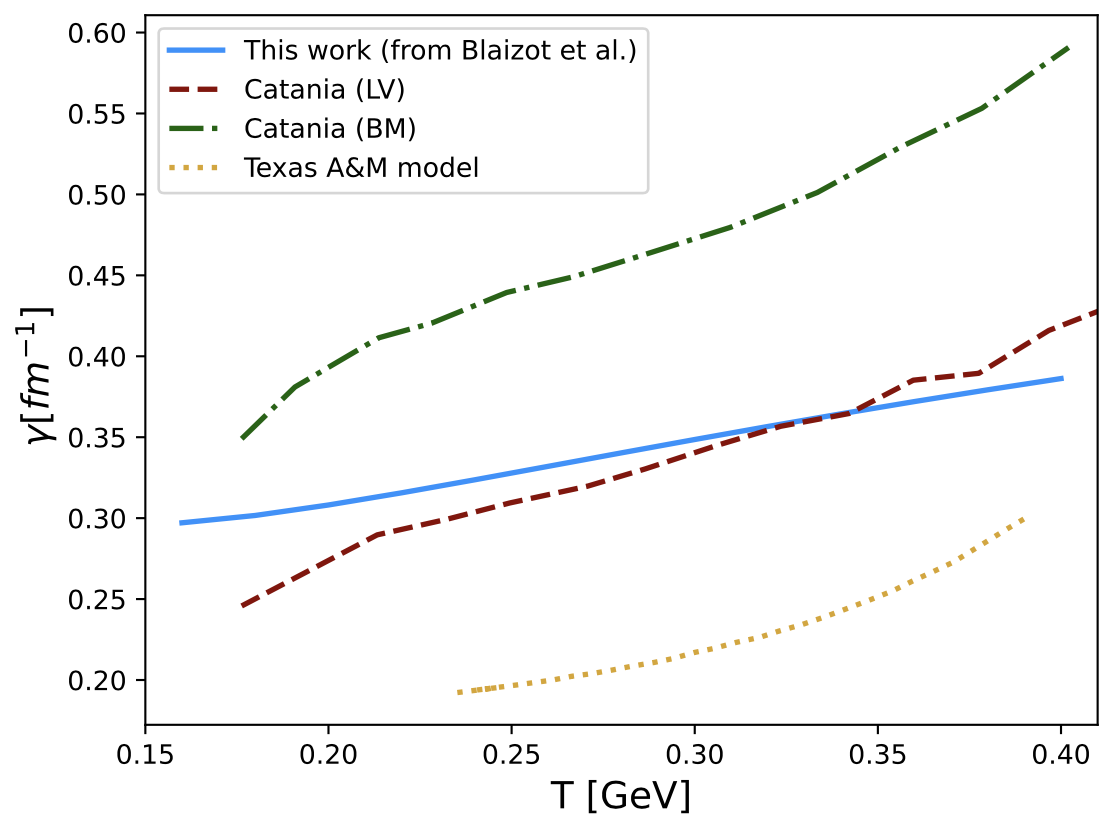}
\includegraphics[width=7.7cm]{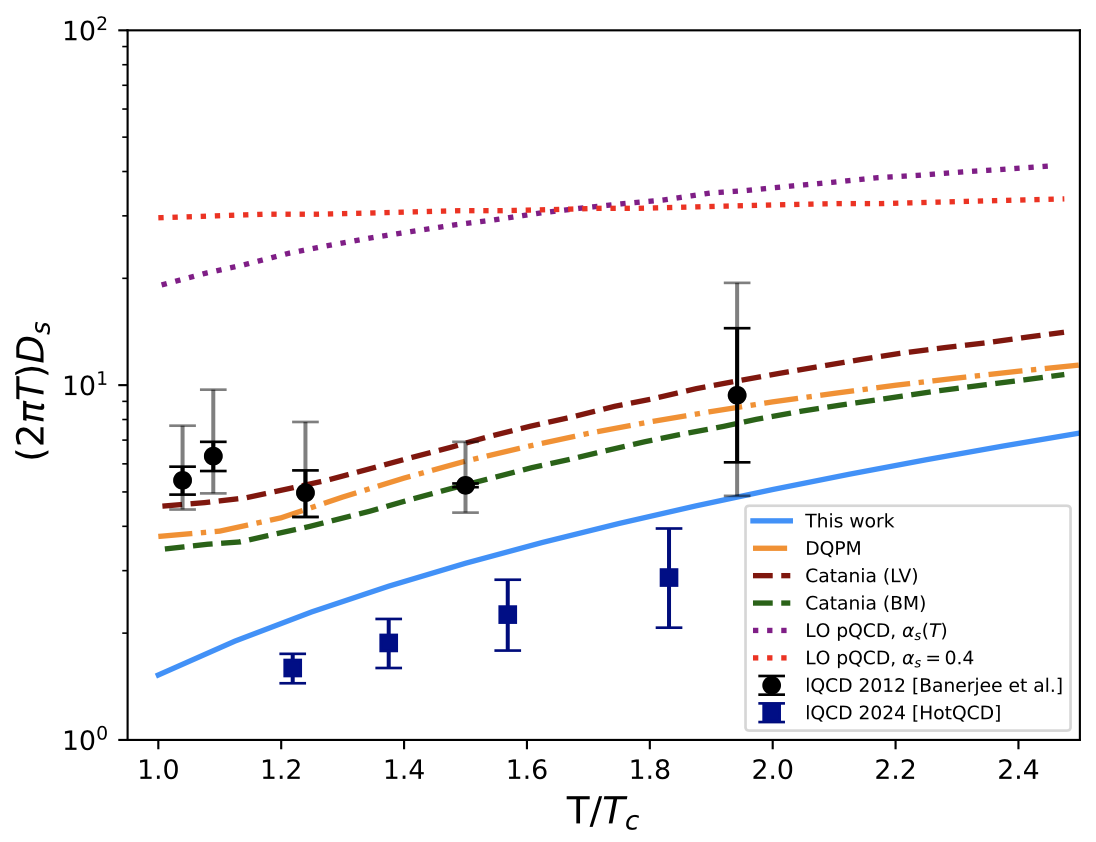}

\caption{Upper panel: Drag coefficient as a function of temperature compared to results from the Texas A\&M model~\cite{Huggins:2012dj} and the Catania quasi-particle model~\cite{Scardina:2017ipo}. Lower panel: Spatial diffusion coefficient of the charm quarks as a function of the temperature ($T_c = 160 \ \text{MeV}$) compared to the same models and to results from the Dynamical Quasi-Particle Model~\cite{Song:2024hvv}, as well as data from lattice QCD~\cite{Banerjee_2012, Altenkort_2024} and pQCD calculations~\cite{Moore_2005,van_Hees_2005}.}
\label{fig:drag_coefficient}
\end{figure}
% Notably the purple dotted line in fig.. differs significantly from the blue line. This is related to the difference in charm mass and coupling constant used for both of these calculations. 

The left panel of Fig.~\ref{fig:drag_coefficient} shows the drag coefficient $\gamma$ for different temperatures, illustrating an enhancement with increasing temperatures. Compared with other heavy-quark drag force determinations in the literature, our values are in the same physical range~\cite{Rapp:2018qla}. In this figure specifically we compare the drag coefficient resulting from the employed potential model with the results of the Texas A\&M model~\cite{Huggins:2012dj} based on a $T$-matrix calculation using the heavy-quark internal energy from lattice-QCD as a potential, and the Catania quasi-particle model~\cite{Scardina:2017ipo} extracted from simulations based on Langevin (LV) or Boltzmann (BM) equations. In the lower, right panel the spatial diffusion coefficient $D_s$ of the charm quarks is displayed, which is directly related to the mean-square displacement according to
\begin{equation}
    \langle \bm{r}^2(t) \rangle = 6 D_s t\ .
\end{equation}
Hence, $D_s$ quantifies how quickly the charm quarks spread out spatially over time due to random thermal fluctuations and interactions with the medium and is connected with the drag coefficient via
\begin{equation}
    D_s = \frac{T}{M_c \gamma}\ . \label{eq:Einstein}
\end{equation}
The spatial diffusion coefficient in our simulation, which results from the drag force proposed in~\cite{Blaizot}, appears to be smaller than the models we are comparing to, while the most recent data from lattice QCD~\cite{Altenkort_2024} supports smaller values of $D_s$. Possible reasons for the small magnitudes of $D_s$ is the application of a stronger coupling $\alpha_s$, as mentioned above, as well as using a charm-quark mass of $m_c = 1.8\ \text{GeV}$, as opposed to the pQCD calculations~\cite{Moore_2005,van_Hees_2005} which were obtained using $M_c = 1.4\  \text{GeV}$ and $M_c = 1.5\  \text{GeV}$ respectively.

We note that the results for $D_s$ from the Catania group are not calculated using the drag force~(\ref{eq:Einstein}) like in our case, but in an independent way using $D_s=T^2/D(0)$, with the diffusion coefficient in momentum space $D(p)$ of Eq.~(\ref{eq:CD}). The differences between the two methods are described in Ref.~\cite{Das:2016llg}.

%but still lying within the uncertainties of the data from lattice QCD. 

Adding the force term resulting from the potential (\ref{eq:Blaizot})---allowing for the formation of bound states---the Langevin update steps for a heavy quark and a heavy antiquark correspondingly read,
% \begin{empheq}[left=\empheqlbrace]{align}
%     \bm{r}_{i+\frac{1}{2}} & = \bm{r}_{i} + \frac{\bm{p}_{i}}{E(p)} \frac{\Delta t}{2} \ ,  \\
%     \bm{p}_{i+1} & = \bm{p}_{i} - \bm{F}(\bm{r}_{i+\frac{1}{2}}, \bar{\bm{r}}_{i+\frac{1}{2}}) \Delta t \label{mom_update} \\
%     & - \gamma \bm{p}_{i} \Delta t + \sqrt{2ET \gamma \Delta t } \bm{\rho}
%      \ , \nonumber \\
%     \bm{r}_{i+1} & = \bm{r}_{i+\frac{1}{2}} + \frac{\bm{p}_{i+1}}{E(p)} \frac{\Delta t}{2} \ ,
% \end{empheq}

\begin{empheq}[left=\empheqlbrace]{align}
    \bm{r}_{i+\frac{1}{2}}^\alpha & = \bm{r}_{i}^\alpha + \frac{\bm{p}_{i}^\alpha}{E(p^\alpha)} \frac{\Delta t}{2} \ ,  \\
    \bm{p}_{i+1}^\alpha & = \bm{p}_{i}^\alpha - \bm{F}(\bm{r}_{i+\frac{1}{2}}^\alpha, \bar{\bm{r}}_{i+\frac{1}{2}}^\alpha) \Delta t \\
    & - \gamma \bm{p}_{i}^\alpha \Delta t + \sqrt{2ET \gamma \Delta t } \bm{\rho}_\alpha \ , \nonumber \\
    \bm{r}_{i+1}^\alpha & = \bm{r}_{i+\frac{1}{2}}^\alpha + \frac{\bm{p}_{i+1}^\alpha}{E(p^\alpha)} \frac{\Delta t}{2} \ ,
    \label{updatestep3}
\end{empheq}
where $i$ denotes a time step $dt$, \( \alpha \in \{ Q, \bar{Q} \} \) and analogous for multiple heavy-quark pairs. Using these expressions, the dynamics of the selected number of charm-anticharm quark pairs are simulated, where the quark and antiquark always interact pairwise. \\
With this prescription one can now analyze in detail the microscopic formation, dissociation, the dynamical equilibration and the thermal population of charmonium states. We note that such a microscopic model has already been proposed originally by Shuryak and Young~\cite{Young:2008he,Young:2009tj} and also in~\cite{Blaizot}.

\section{Dynamical Bound State Formation}
\label{sec:BSformation}

In order to identify a charmonium state we use a classical definition in which a charm-anticharm pair is bound if the relative energy of the pair is smaller than zero. This binding energy is calculated by subtracting the total energy of the pair to the sum of the individual charm and anticharm quarks as well as the potential acting between them,
\begin{align}
\begin{split}
E_{c \, \bar{c}} = E_c + E_{\bar{c}} + V(|\bm{r}_c - \bm{r}_{\bar{c}}|) - E_{\text{tot}} \\
= \sqrt{m_c^2 + \bm{p}_c^2} + \sqrt{m_{\bar{c}}^2 + \bm{p}_{\bar{c}}^2} + V(|\bm{r}_c - \bm{r}_{\bar{c}}|)  \\
+ \sqrt{(m_c + m_{\bar{c}})^2 + (\bm{p}_c + \bm{p}_{\bar{c}})^2}  \, .
\end{split}
\label{e_rel}
\end{align}

\begin{figure}[h]
\centering
\includegraphics[width=8cm]{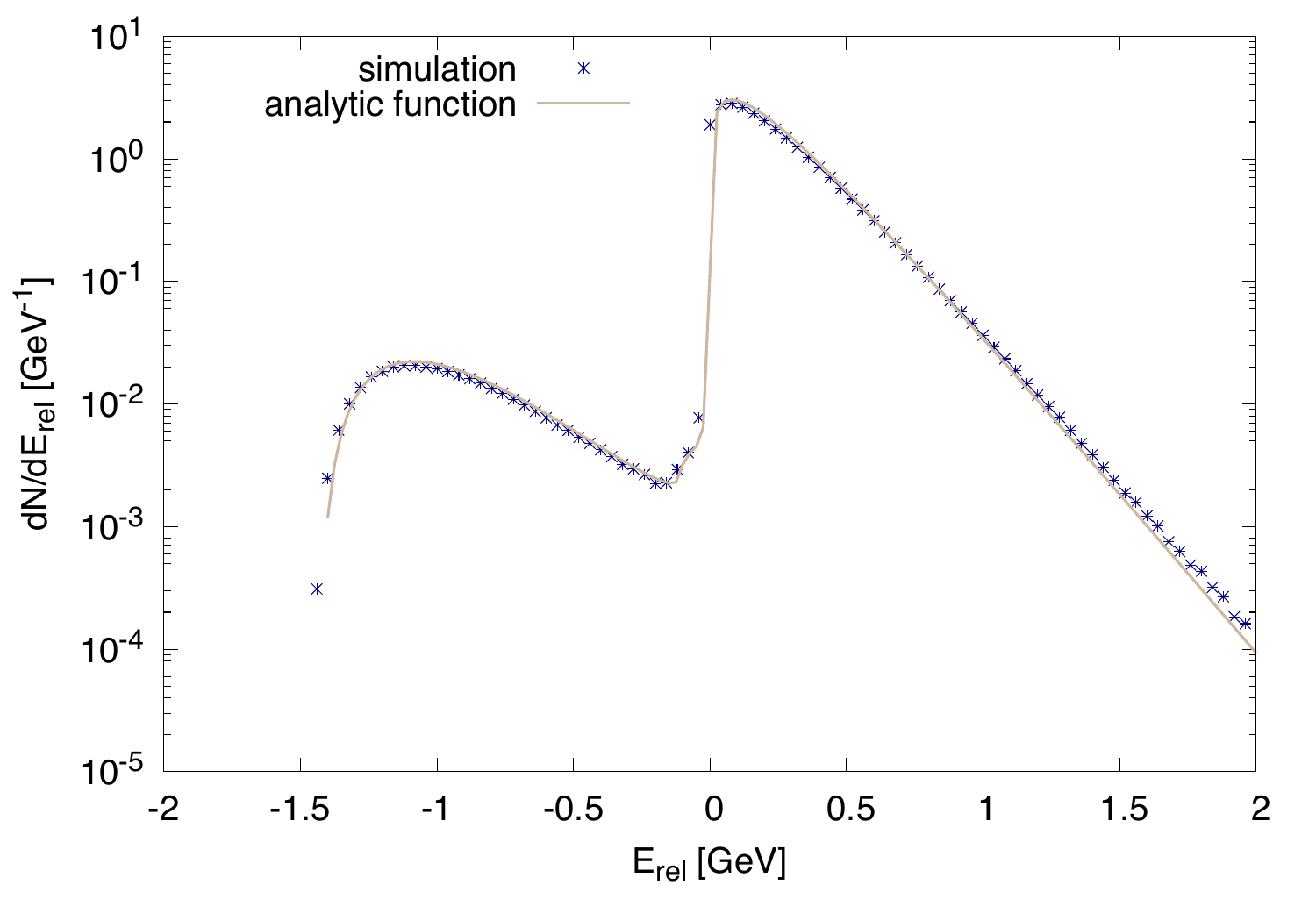}
\includegraphics[width=7.7cm]{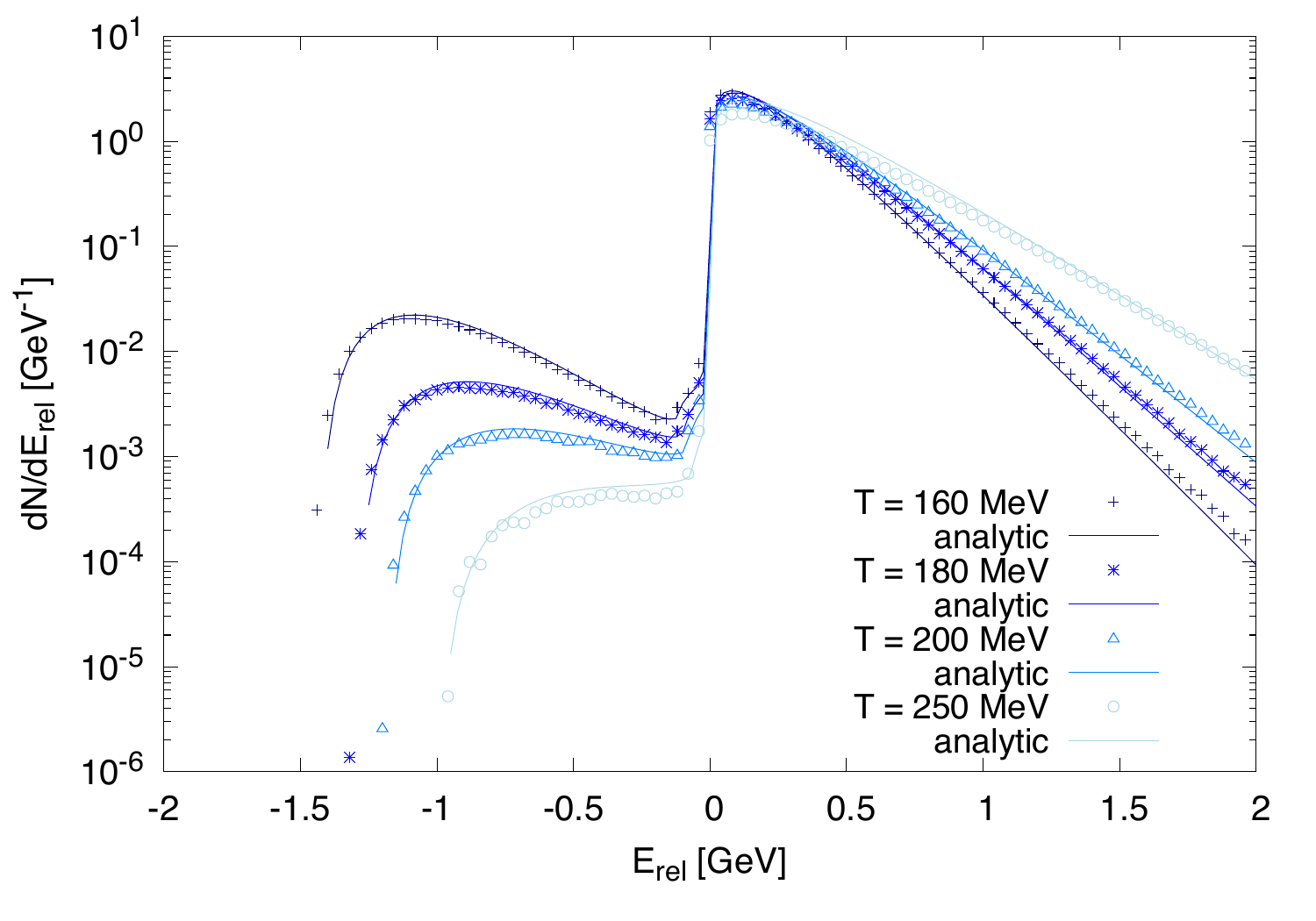}
\caption{Upper panel: Simulation of the distribution of the relative energy of the heavy-quark pair in a cubic box with side length $10\ \text{fm}$ at $T=160\ \text{MeV}$ compared the analytic expectation~(\ref{dN_dErel}). Lower panel: Comparison of the results of the simulation of the relative energy for different temperatures.}
\label{fig:ebin}
\end{figure}

This is checked at every time step, and for every pair combination. Once the system has thermalized, the distribution of the relative energy is expected to be given by the classical density of states, multiplied with the Boltzmann factor in order to obtain the thermal distribution of states, 
\begin{align}
\frac{dN_{c \bar{c}}}{dE_{ c \bar{c}}} &= g(E_{c \bar{c}}) \exp{\left(- \frac{E_{c \bar{c}}}{T}\right)} \nonumber \\
&= \int \frac{d^3r d^3p}{(2\pi \hbar)^3} \delta \left( E- \frac{p_{rel}^2}{2 \mu}- V(r) \right) \exp{\left(- \frac{E_{c \bar{c}}}{T}\right)} \nonumber \\
&= (4\pi)^2 (2 \mu)^{\frac{3}{2}} C
    \int_0^R dr r^2 \exp \left(-\frac{E_{c \bar{c}}}{T}\right) \sqrt{E_{c \bar{c}}-V(r)} \ ,    \label{dN_dErel}
\end{align}
where $\mu=M_c/2$, $C$ is a normalization constant (fixed to the total number of pairs $N_{\rm{pair}}$) and $V(r)=\textrm{Re } \mathcal{V}(r)$, cf. Fig.~\ref{fig:potential}. The upper boundary of the integral $R$ corresponds to the radius of a sphere with the same volume as the box of our simulation. 
In order to investigate whether our description leads to the right equilibrium density of states, we perform box simulations ($N_{\rm{event}}=4\cdot 10^4$) with a single heavy-quark pair ($N_{{\rm pair}}=1$) at a constant temperature and volume. The heavy quarks follow the Langevin equation with $T=160$ MeV in a cubic box of volume $V=(8 \ \text{fm})^3$ and periodic boundary conditions. Figure~\ref{fig:ebin} shows the comparison between the simulation in the long-time limit, $t=400\ \text{fm}$, and the theoretical expectation of Eq.~(\ref{dN_dErel}), which demonstrates a nearly perfect agreement: The first bump in the left side of the plot corresponds to bound states, where the relative energy of the charm-anticharm-quark pair is negative. Clearly, only a small fraction of bound states occurs, with the strongest bound state at approximately $1 \ \text{GeV}$ of binding energy. Accordingly, the right side peak, with positive values of the relative energy, display the distribution of free charm and anticharm quarks. Consequently, a charmonium state is identified as a state in the negative energy side of the distribution. The lower panel of Fig.~\ref{fig:ebin} compares the distribution of the relative energy of simulations with different temperatures of the system. Here once again the screening of the potential becomes apparent as the bump on the left side corresponding to the bound states decreases with increasing temperatures. Therefore, potential bound states become less likely, as expected.

\begin{figure}[h]
    \centering
    \includegraphics[width=8cm]{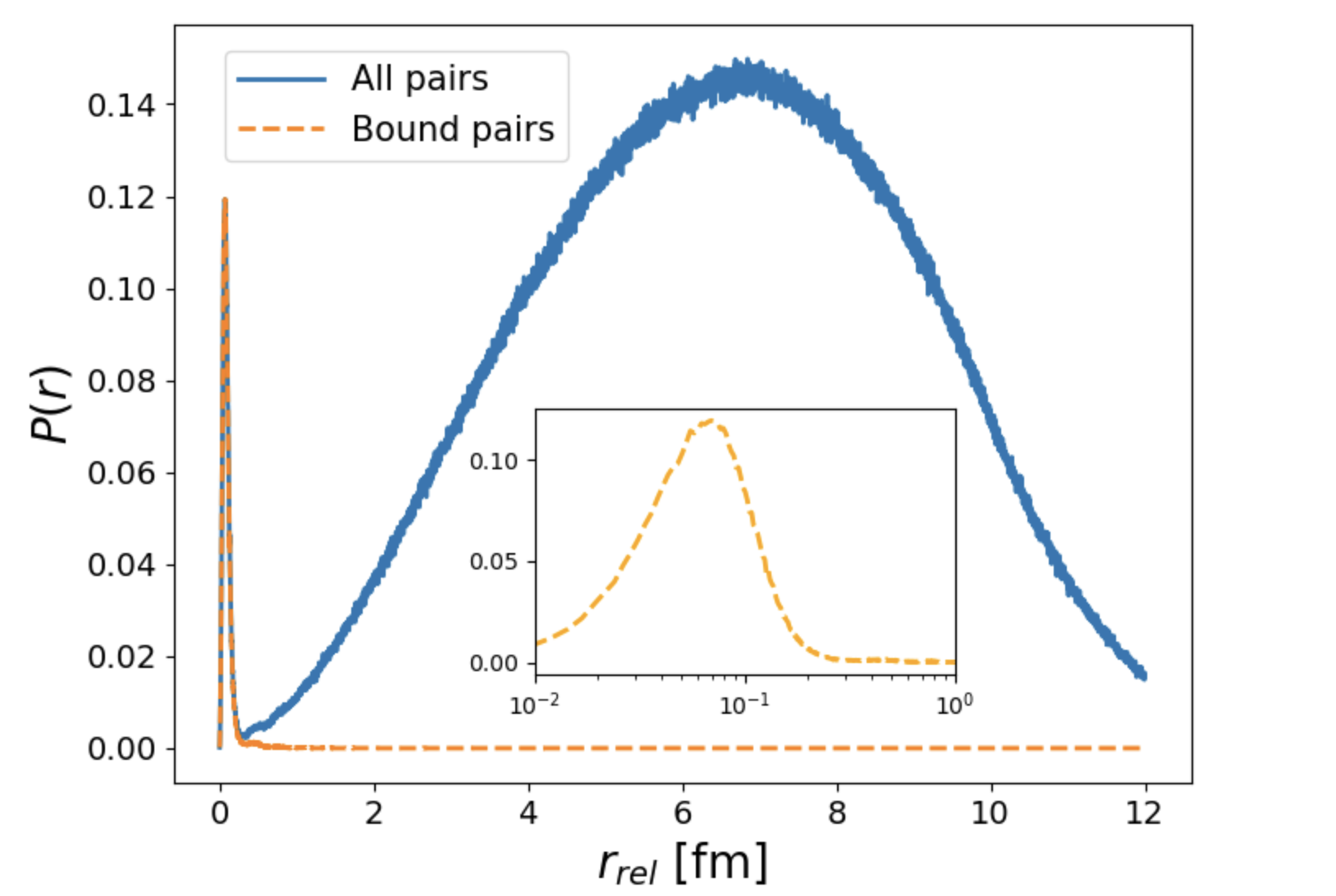}
    \caption{The probability distribution of the relative distance between a charm-anticharm quark pair, calculated with the same parameters as in Fig.~\ref{fig:ebin}. The blue line depicts the distribution for all pairs, regardless of their relative energy, the orange dashed line corresponds to bound states, i.e. pairs with a negative relative energy.}
    \label{fig:distance-dist}
\end{figure}

In order to confirm that Eq.(\ref{e_rel}) is a suitable criterion for identifying charmonium states, we consider the probability distribution of the relative distance of the charm-anticharm quark pairs. In Fig.~\ref{fig:distance-dist} we compare the distribution of all pairs, regardless of their relative energy (blue line) with the case where only pairs with a negative relative energy are included (dashed orange line). The comparison reveals that in both cases, a narrow peak at small distances emerges, corresponding to the relative distance between a quark and antiquark within a bound state. At larger distances, the potential between the heavy quarks becomes negligible, leading to a distance distribution depending on the box size, which is represented by the second peak of the blue line. Since this second peak is absent when only considering pairs with a negative relative energy, we conclude that Eq.~(\ref{e_rel}) is a suitable benchmark for only pairs with a reasonably small relative distance as bound states with within our model.

When more than one heavy-quark pair is present in the system, it is possible for a single quark to have a negative relative energy with more than one antiquark. Since we do not allow for more than two charm quarks to bind, we need to ensure that one quark can only form a bound state with one antiquark at a certain time step. Therefore, the relative energy between every quark and antiquark is calculated at a time step and stored. The pair with the lowest binding energy is selected as the first bound state and all other entries containing one of the partners are deleted. Now, the pair with the next lower energy is identified and selected. This procedure is continued until the following entry is one with a positive binding energy. In that case, all bound states at the respective time step have been found and identified, and the simulation continues to the next time step. In this way, we can monitor the number of bound states along the time evolution of the simulation. We perform this study in a somewhat larger box (since we will also add a higher number of heavy-quark pairs, $N_{\rm{pair}}$) with volume of $V=(10\ \text{fm})^3$ at $T = 160\ \text{MeV}$.

\begin{figure}[h]
    \centering
    \includegraphics[width=8cm]{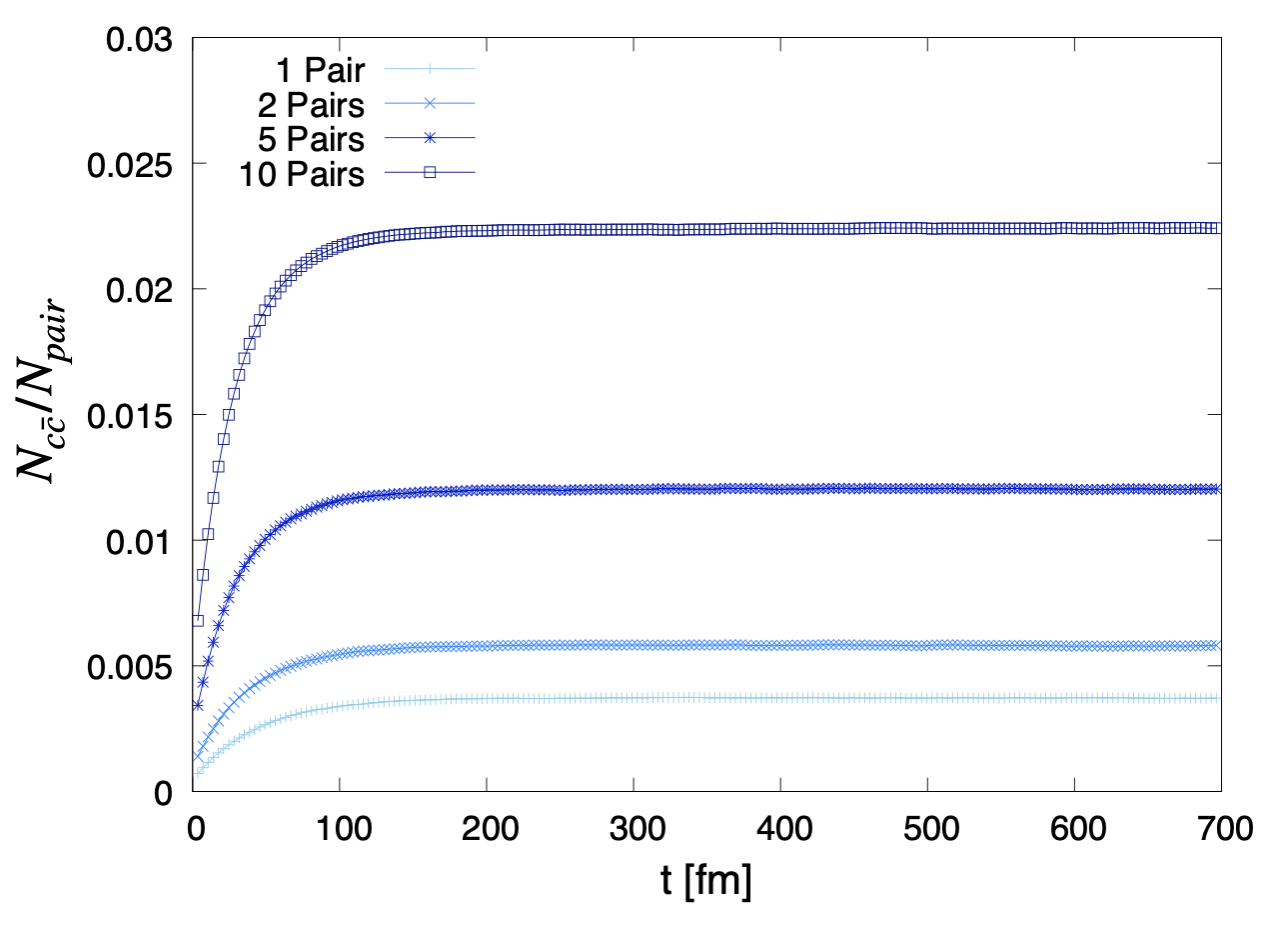}
    \caption{Time evolution of the fraction of bound states in box simulations with $N_{\rm{pair}}=1, 2, 5 \text{ and}\ 10$ initial heavy-quark pairs, respectively at constant temperature $T=180\ \text{MeV}$ and volume $V=(10\ \text{fm})^3$.}
    \label{fig:time_evol}
\end{figure}

The heavy quarks are placed randomly inside the cubic box, therefore there is a chance that two of them are placed close enough together in order to form a bound state directly at $t=0$, which is reflected by the non-zero initial values of each curve of Fig.~\ref{fig:time_evol}. With increasing time, recombination and dissociation processes take place until the potential equilibrium limit is reached, where the number of dissociation is equal to the number of recombination processes. When a larger number of pairs are present in the box, the probability to form a bound state increases due to the higher density of heavy quarks, which becomes apparent in Fig.~\ref{fig:time_evol} in the fact that the equilibrium value of the fraction of bound states increases with the number of pairs, i.e. $N_{J/\psi} \propto N_c^2 \cdot \frac{1}{V}$. This dependence will be analyzed in more detail later.

\begin{figure}[h]
\centering
\includegraphics[width=8cm]{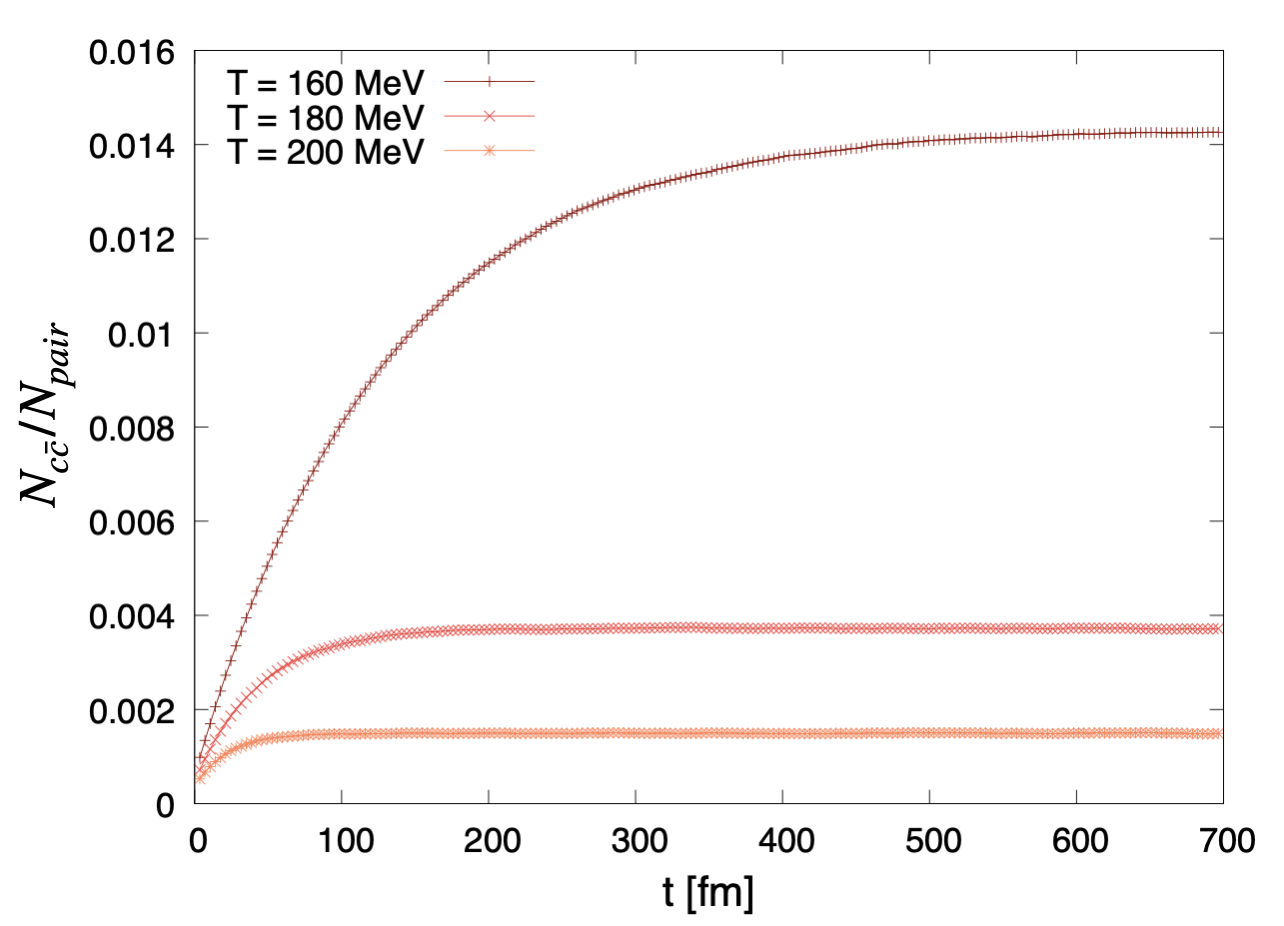}
\includegraphics[width=7.7cm]{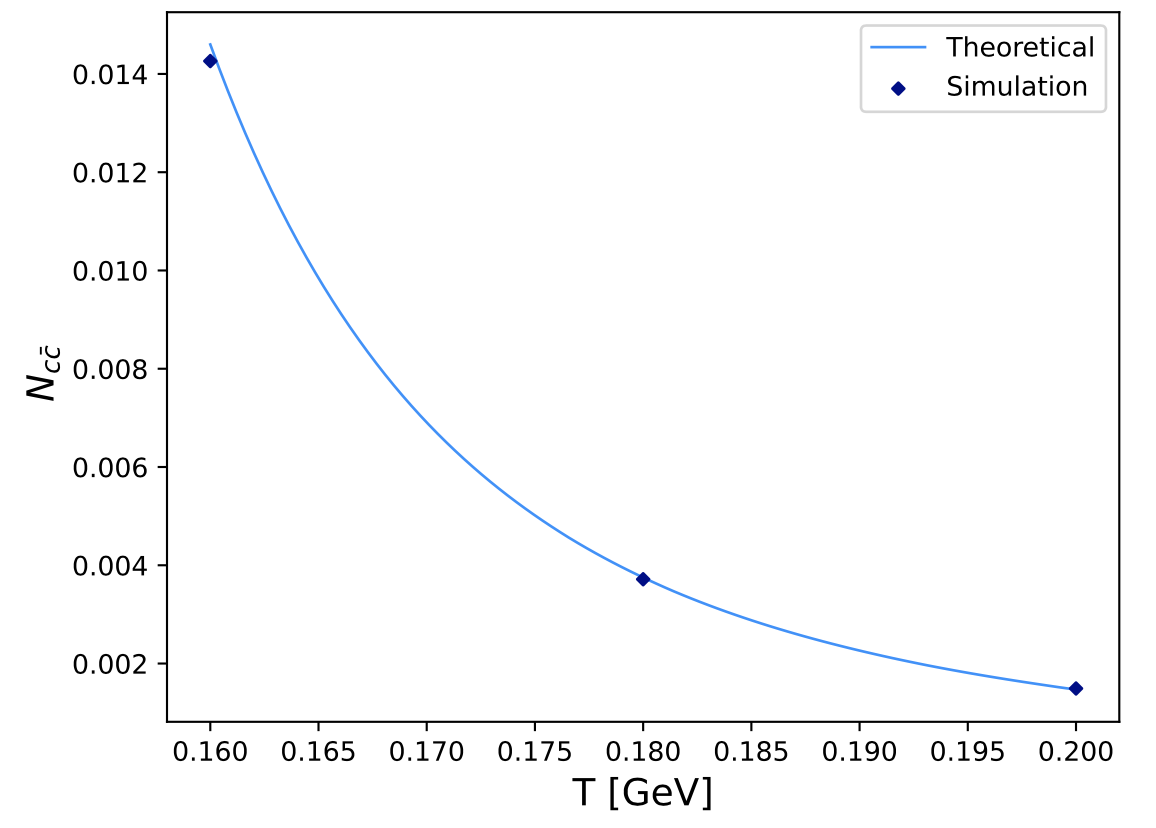}
\caption{Temperature dependence of the formation of charmonium bound states in a box calculation at constant volume of $V = (10\ \text{fm})^3$ and $N_{\rm{pair}}=1$. Upper panel: Fraction of bound states as a function of time for three different temperature. Lower panel: Equilibrium charmonium fraction as a function of temperature. Results from numerical simulations versus theoretical expectation from Eq.~(\ref{eq:theocharmyield}).}
\label{fig:T-Boundstates}
\end{figure}

As a next analysis, the dependence of the temperature of the system on the formation of bound states can be investigated. In the upper panel of Fig.~\ref{fig:T-Boundstates} the time evolution of the bound-state formation is depicted for three different temperatures. It can be directly seen from the graph that at higher temperatures the formation of bound states is suppressed due to the screening of the potential. Correspondingly, the highest charmonium yield is obtained at $T = 160\ \text{MeV}$. Furthermore, the scaling of the charmonium yield with the medium temperature is summarized in the right panel of Fig.~\ref{fig:T-Boundstates}, where we show the numerical result of the simulation for three temperatures, $T=160, 180, 200$ MeV, together with the theoretical equilibrium multiplicities coming from the energy integration of Eq.~(\ref{dN_dErel}),
\begin{equation}
    N_{c\bar{c}} = \int_{E_{c \bar{c},\rm{min}}}^0 dE_{c \bar{c}}  \ \frac{ dN_{c \bar{c}}}{dE_{c \bar{c}}} \ . \label{eq:theocharmyield}
\end{equation}
We find an excellent agreement between the theoretical expectation and the simulations.

% \begin{figure}[h]
%     \centering
%     \includegraphics[width=8cm]{plots/bc_temperature.pdf}
%     \caption{The time evolution of the fraction of bound states for various temperatures at a constant volume of $V = 8^3\ \text{fm}^3$ within box simulations with a single charm-anticharm pair}
%     \label{fig:bc_T}
% \end{figure}

Finally, the effect of the system's size on the regeneration probability can by studied by varying the side length of the box.

\begin{figure}[h]
    \centering
    \includegraphics[width=8cm]{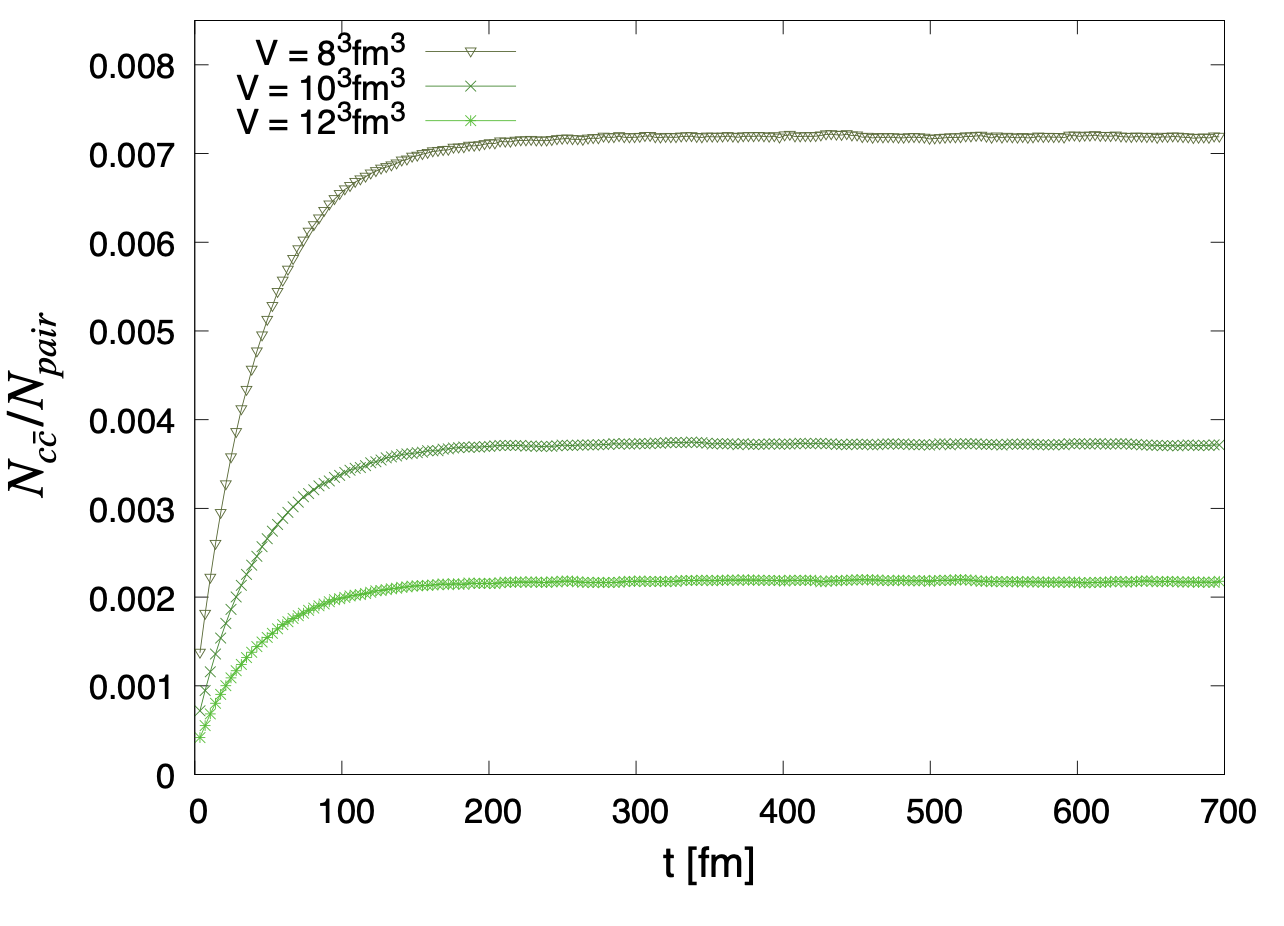}
    \caption{The time evolution of the fraction of bound states for different box volumes at a constant temperature of $T = 180 \text{ MeV}$ in simulations with a single charm-anticharm pair.}
    \label{fig:bc_V}
\end{figure}

Figure~\ref{fig:bc_V} shows the time evolution of the fraction of bound states for boxes with side lengths $8, 10$ and $12\ \text{fm}$ respectively. As expected, the equilibrium value of the fraction of bound states is higher for a smaller box volume, since the probability for a charm to encounter an anticharm quark and therefore interact to form a bound state is enhanced in a reduced system size.

\begin{figure}[h]
    \centering
    \includegraphics[width=8cm]{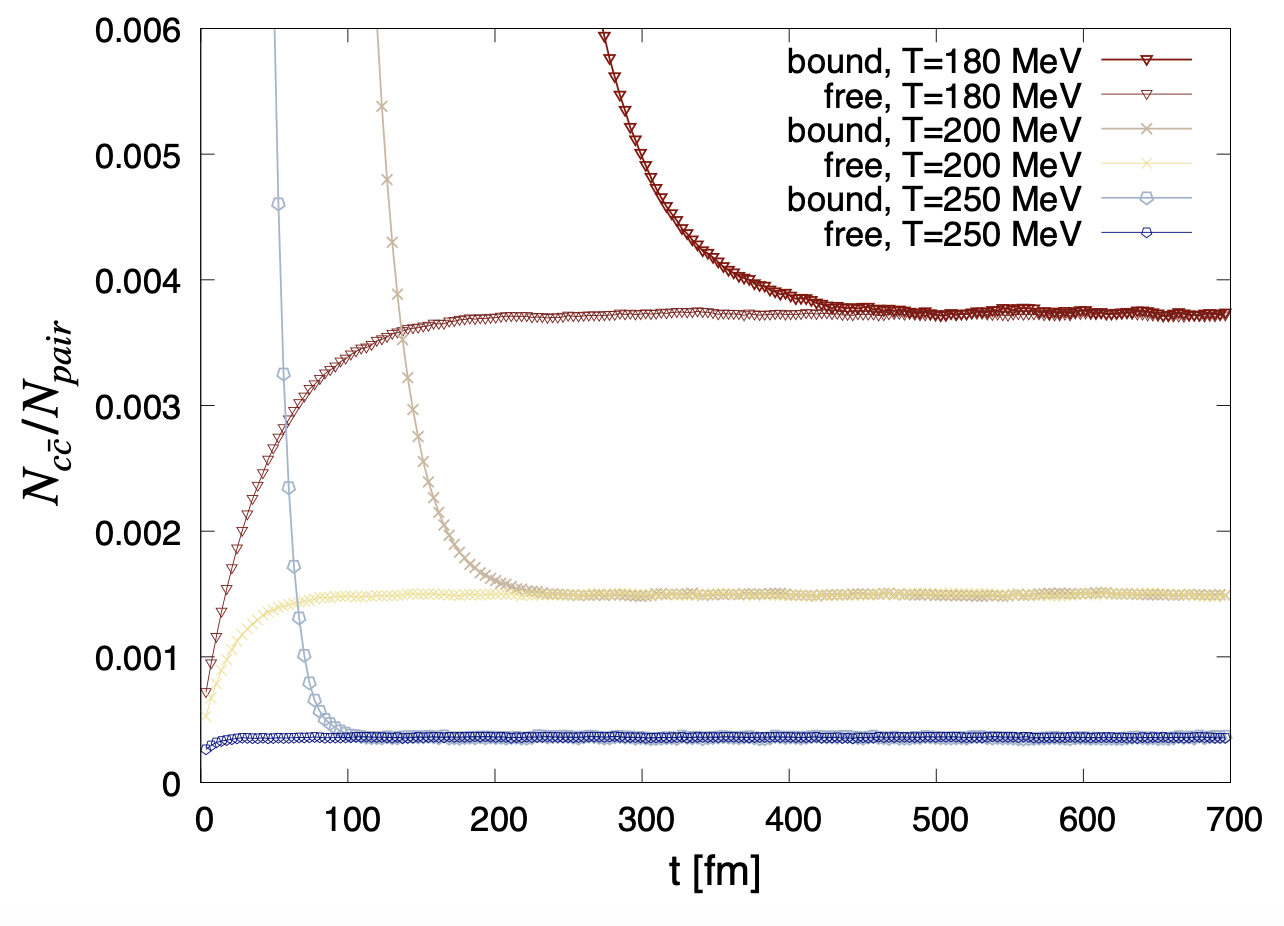}
    \caption{Comparison of the bound-state time evolution of initially free and bound pairs for different temperatures with a single charm-anticharm pair in a box of volume $V=(10\ \text{fm})^3$.}
    \label{fig:free-vs-bound}
\end{figure}

Finally, we can demonstrate that the same equilibrium charmonium yield is obtained, regardless whether the charm-anticharm pair is initially bound or free. Fig.~\ref{fig:free-vs-bound} shows the time evolution of the fraction of bound states for various temperatures for both cases. In the case of initially bound pairs the time to reach the equilibrium limit is larger than for the free quarks, but for all three considered temperatures the same value is reached in the long-time limit, assuring that dissociation and regeneration processes occur in a detailed balance and demonstrating the thermal equilibration of bound states.
 
\section{Momentum relaxation and chemical equilibration time}
\label{sec:eq_time}

\begin{figure}[h]
    \centering
    \includegraphics[width=8cm]{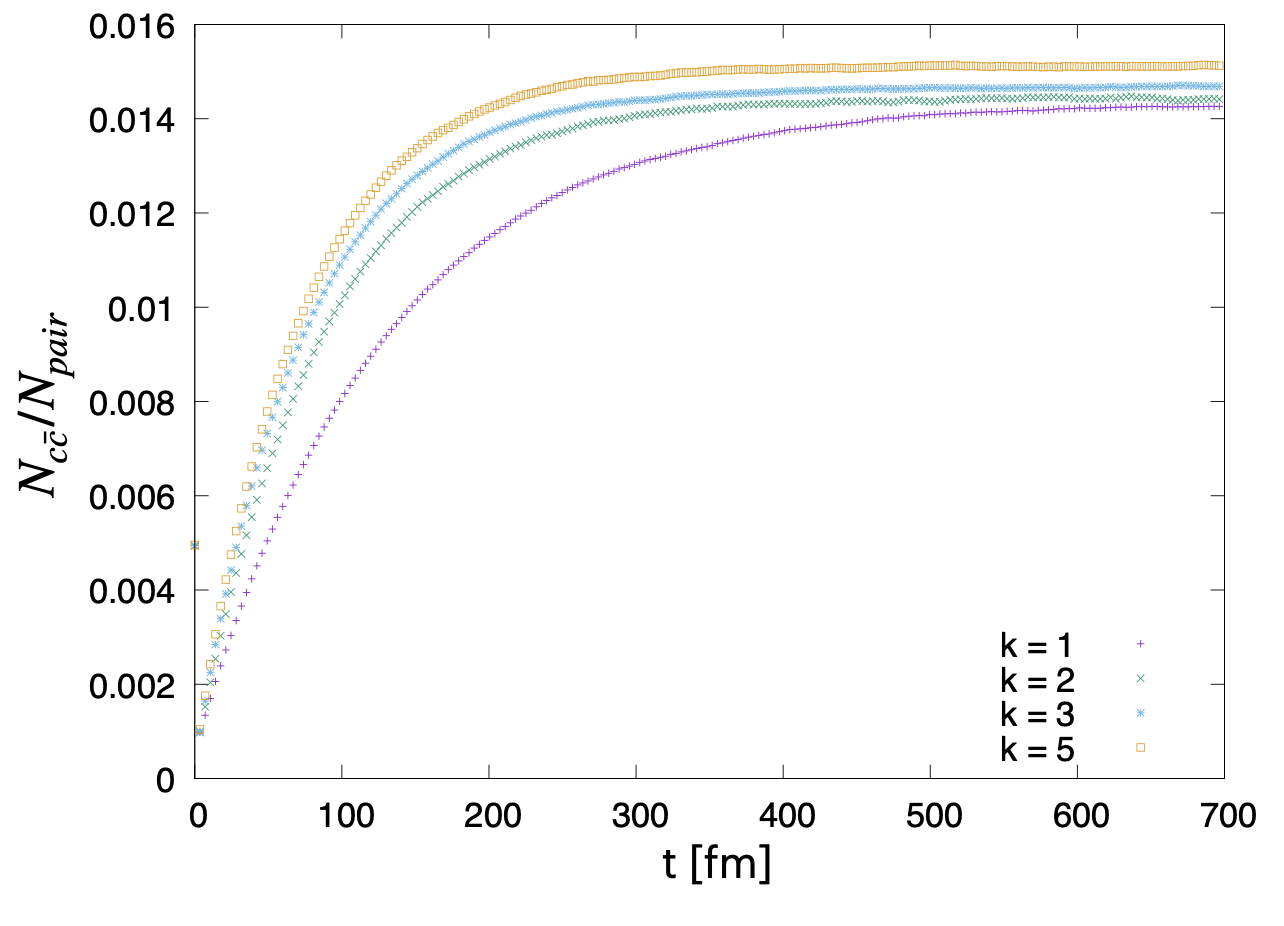}
    \caption{Time evolution of the fraction of bound states for different scaling of the drag coefficient $k\gamma$ within box simulations with a single charm-anticharm pair, at constant temperature $T=160\ \text{MeV}$ and volume $V=(10\ \text{fm})^3$.}
    \label{fig:bc_drag}
\end{figure}

As mentioned before, the drag force quantifies the amount of interactions of the heavy quarks with the surrounding medium. Therefore it determines the relaxation rate of the average particle's momentum to its equilibrium value. The chemical equilibration time of the system---to form the final charmonia states--- should depend inversely on the drag coefficient in the Langevin equation and is of the same order as $\tau_R$. In order to study the influence of the drag coefficient on the equilibration time, different simulations with a single charm-anticharm pair are run with the parameter $\gamma$ scaled up by a factor $k=2,3$ and $5$. The results for the situation of one charm-quark pair are depicted in Fig.~\ref{fig:bc_drag}. 

From the analysis of the curves it can be concluded that a higher drag coefficient leads to a somewhat faster chemical equilibration of the system. This might be expected since to achieve equilibration, particles need to relax their momentum first, which they do in time scales of the order of the inverse of the drag coefficient, $\tau_R=\gamma^{-1}$. However, full chemical equilibration---understood as achieving a time-independent number of bound states---is dominated by a much slower process, the genuine two-body dynamics that makes the eventual generation of charmonia, and dominated by the time scales of the pairwise potential $V(r)$. Both effects, momentum relaxation and chemical equilibration, are account in the Langevin equation. To quantify the latter, we make an exponential ansatz of the form,
\begin{equation}
    N_{c \bar{c} } (t)= N_{c \bar{c} ,\rm{eq}} \left( 1- A e^{ -t /\tau_{\rm{eq}}} \right) \ , 
\end{equation}
to the obtained time evolution curves of Fig.~\ref{fig:bc_drag}. The resulting chemical equilibration times---together with their respective relaxation times---are shown in Table~\ref{tab:times}.
\begin{table}[h]
\begin{tabular}{|c|c|c|}
\hline
$k$ & $\tau_R$ (fm) & $\tau_{\rm{eq}}$ (fm) \\ 
\hline
1 & 3.3 & 133.78 \\ %127.3
2 & 1.7 & 89.31 \\ %89.306 %84.1
3 & 1.1 & 78.89  \\ %78.886 %74.0
5 & 0.7 & 70.75 \\ %70.8
\hline
\end{tabular}
\caption{Comparison of the relaxation and chemical equilibration times for each of the curves in Fig.~\ref{fig:bc_drag} with different scaling factor of the drag force coefficient ($k\gamma$) used in the simulation at $T=160$ MeV.}
\label{tab:times}
\end{table}
From the values of this table it is clear that there is separation of time scales in the heavy-quark thermalization. While the relaxation time (given by the inverse drag coefficient) is rather fast, the final pair formation takes much more time to equilibrate towards the final charmonium multiplicities. However, one cannot naively conclude that this comes from a very distinct time scales in the imaginary and real parts of the potential since, as we will see, the equilibration rate depends strongly on temperature. To furthermore quantity this effect, the calculation can be repeated for various temperature. The results are summarized in Table~\ref{tab:times2}. The general trend is similar as before but we also notice the fast decrease of the equilibration time towards higher temperatures and the large difference in time scales between $T=160\ \text{MeV}$ and higher temperatures. This reflects the strong binding force of the potential at this particular temperature, as depicted in Fig.~\ref{fig:potential}. At low temperatures, it is difficult for a bound pair to escape from the bottom of a deeper potential with random kicks that are also of less intensity than at high temperatures, where the potential is also less bound and pair melting is more favorable. For the regeneration of bound states, the strength of random kicks is larger at higher temperature and that helps to bring the two particles together, while at low temperatures the noise is much weaker and the pair needs more time to roll down the potential.

\begin{table}[h!]
\centering
\begin{tabular}{|c|c|c|c|c|}
\hline
\multirow{2}{*}{$k$} & \multicolumn{4}{c|}{ $\tau_{\rm{eq}}$ (fm)} \\ \cline{2-5} 
                   & \(T = 160\) MeV & \(180\) MeV & \(200\) MeV & \(250\) MeV \\ \hline
1                  & 133.78      & 43.88      & 22.25      & 5.87       \\ \hline
%1                  & 133.78      & 43.878      & 22.252      & 5.873       \\ \hline
2                  & 89.31      & 30.60      & 14.22      & 4.88        \\ \hline
%2                  & 89.306      & 30.604      & 14.221      & 4.88        \\ \hline
3                  & 78.89      & 28.82      & 13.68      & 4.80       \\ \hline
%3                  & 78.886      & 28.817      & 13.685      & 4.797       \\ \hline
5                  & 70.75       & 24.42      & 9.78       & 4.37       \\ 
%5                  & 70.75       & 24.423      & 9.781       & 4.366       \\ 
\hline
\end{tabular}
\caption{The relaxation time for different scaling coefficients of the drag coefficient $k\gamma$ at different temperatures.} \label{tab:times2}
\end{table}

\section{Equilibrium Yields and Comparison to Statistical Hadronization Model}
\label{sec:sectionSHM}

To investigate further whether our results are in accordance with a thermodynamically equilibrated system, the final charmonium yield in equilibrium can be compared to the SHM~\cite{Andronic:2017pug}, where
statistically equilibrated source volume can be described in terms of a grand-canonical ensemble (GCE). Therefore, we can verify if our results can be made consistent with the SHM by calculating the number of charmonium states in the GCE. Apart from the charmonium ground state $J/\psi$ we take into account further charmonium states, namely $\eta_c$, $\chi_c$ and $\psi(2S)$, since one cannot distinguish one from another in the classical simulation, and only a continuum of bound states are generated dynamically.
The average particle number of a certain species in the GCE can be calculated using the quantum mechanical partition function according to,
\begin{equation}
    N = T \left( \frac{\partial \ln {\cal Z}}{\partial \mu} \right)_T \ ,
\end{equation}
with 
\begin{equation}
    \ln {\cal Z} = a \sum_\alpha \ln \left[ 1 + a \exp \left( -\frac{E_\alpha - \mu_\alpha}{T} \right) \right],
\end{equation}
where $a = \pm 1$, depending on whether the particles are Fermi-Dirac or Bose-Einstein particles; and $\alpha$ sum over the different states of the ensemble. In the non-relativistic and classical limit one has the following expression for the computation of the particle number, i.e. charm and anticharm quarks,
% taylor expansion forsmall particle numbers, classical particles, neglecting qm effects (only leading order term), with Bessel function in the nonrelativistic limit $m/T \gg 1$
\begin{equation}
    N_c = \lambda_c g_c V \left( \frac{M_cT}{2\pi} \right)^{3/2} e^{-\frac{M_c}{T}} \ ,
    \label{N_c}
\end{equation}
with $\lambda_c$ the fugacity factor $\lambda_c= e^{\mu_c/T}$, $V$ the volume of the system, $g_c$ the particle degeneracy.

Since the number of charm quarks in our simulation is fixed, we can calculate the charm fugacity from this expression. The process of recombination and dissociation of a $J/\psi$-meson, for example, is similar to a reaction in chemical equilibrium,
\begin{equation}
        c + \bar{c} \ (+g) \longleftrightarrow J/\psi \ (+ g) \ ,
\end{equation}
where the gluons represents the random kicks from the system which the charm and anticharm quarks experience. 
In this reaction, the chemical potentials of the reactive and the product are connected through the relation $\mu_{c \bar{c}} = 2\mu_c$, leading to $\lambda_{c \bar{c}} = \lambda_c^2$ for the fugacity.

 Hence, the charmonium multiplicity at a certain temperature and volume can be calculated according to the following expression,

 \begin{equation}
    \begin{split}
    N_{c \, \bar{c}} &= V \sum_\alpha \lambda_c^2 \, g_\alpha \, \left( \frac{M_\alpha \, T}{2 \pi} \right)^{3/2} \exp(-M_\alpha/T) \ , \\
    \alpha &= \{ J/\psi, \eta_c, \chi_c, \psi(2S) \} \ .
    \end{split}
\end{equation}

Inserting Eq.~(\ref{N_c}) for $\lambda_c$ leads to

\begin{equation}
N_{c \, \bar{c}} = \sum_{\alpha} \frac{N_c^2}{V} \frac{g_{\alpha}}{g_{c^2}}
\left( \frac{2 \pi}{T} \right)^{3/2} \frac{M_{\alpha}^{3/2}}{M_c^3} 
e^{\frac{2M_c - m_{\alpha}}{T}} \, 
\end{equation}

with the respective values for the masses and degeneracies of the considered charmonium states from the Particle Data Group~\cite{ParticleDataGroup:2024cfk} (see Table~\ref{charmoniumstates}).

\begin{table}[h]
\begin{tabular}{|c|c|c|}
\hline
 $\alpha$          &  $m_\alpha$ (MeV) & $g_\alpha$    \\
 \hline
\textbf{$J/\psi$}   &           3097       & 3           \\
\textbf{$\eta_c$}    & 2984                 & 1       \\
\textbf{$\chi_{c0}$}    & 3415                 & 1       \\
\textbf{$\psi (2S)$} & 3686                 & 3    \\  
\hline
\end{tabular}
\caption{Table of masses and spin degeneracies of the charmonium states considered for the calculated final charmonium yield. 
Properties taken from Ref.~\cite{ParticleDataGroup:2024cfk}.}
\label{charmoniumstates}
\end{table}

\begin{figure}[h]
    \includegraphics[width=8cm]{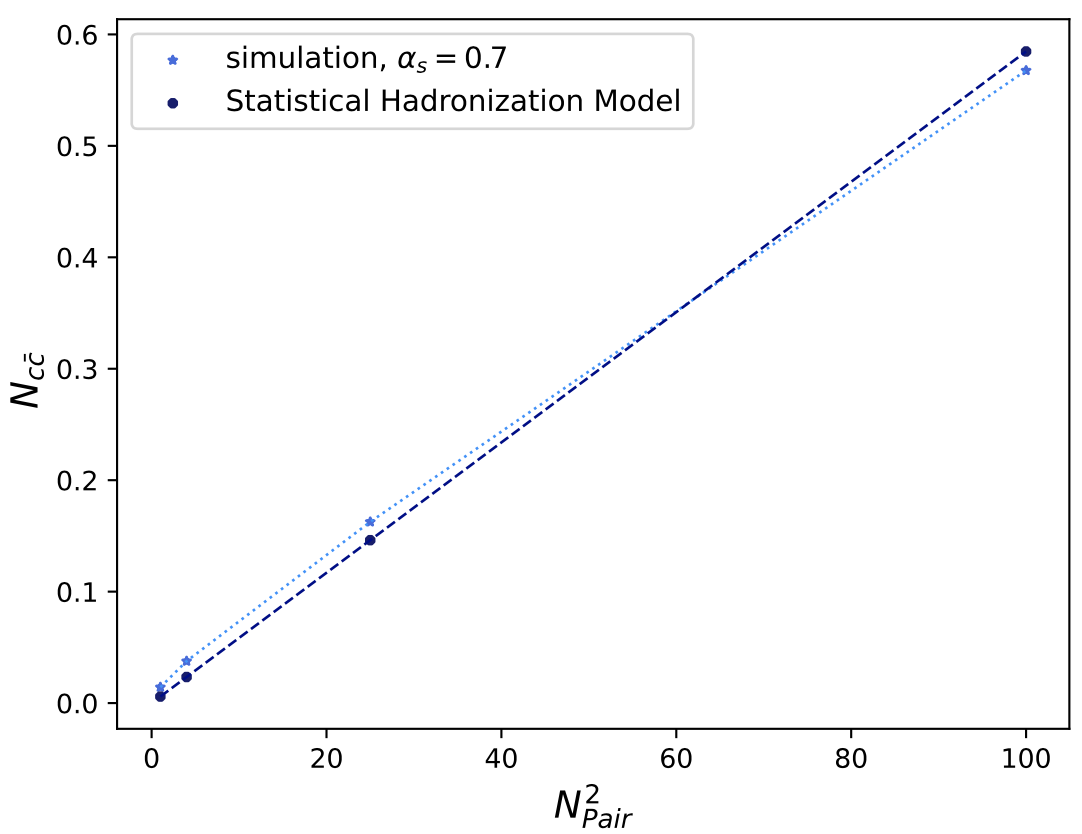}

    \caption{The scaling of the equilibrium charmonium yield with the number of initial pairs squared, compared to predictions from the SHM. Lines are drawn to guide the eye.}
    \label{fig:comparison_SHM}
\end{figure}

In order to set a common baseline with the results from the SHM we need to scale up the strong coupling constant $\alpha_s(T_c)$. This explains why we needed to deviate from the value used in~\cite{Blaizot}. For a larger value of $\alpha_s$ the potential in Fig.~\ref{fig:potential} becomes deeper, therefore an increased formation number of bound states is expected. The chosen value of $\alpha_s = 0.5$ in Ref.~\cite{Blaizot} leads to final multiplicities which would underestimate the yields from the SHM. Hence, one has to choose a slightly larger value of $\alpha_s = 0.7$ in order to match to the SHM. 
In Fig.~\ref{fig:comparison_SHM} as well as in Table~\ref{comparisonSHM} we show the charmonium yields calculated for a box of volume $V = 10^3\ \text{fm}^3$ and $T=160\ \text{MeV} = T_c$ in comparison with the results from our simulation.

\begin{table}[h]
\begin{tabular}{|c|c|c|}
\hline
$N_{\rm{pair}}$ & $N_{c \bar{c}}^{\rm{simulated}}$ & $N_{c \bar{c}}^{\rm{SHM}}$ \\ 
\hline
1              &                                                 0.0143  &  0.0059                                                 \\ 
2              &      0.0376                                             &      0.0236                                          \\
5              &         0.1626                                          &    0.1476                                              \\
10             &              0.5677      &    0.5904        \\         
\hline
\end{tabular}
\caption{Comparison of the final charmonium yield of our simulation with the SHM for different numbers of initial charm-anticharm pairs.}
\label{comparisonSHM}
\end{table}

As can be seen in Fig.~\ref{fig:comparison_SHM} the total charmonium yield scales with the number of charm-anticharm quark pairs squared. This behavior is expected, since every charm-quark can form a charmonium state with every anticharm-quark. \\
With a more closer inspection of Table~\ref{comparisonSHM} one recognizes that for $N_{pair}=1,2$ the simulation manifests slightly larger values than the SHM calculation. It has to be noted that the gauging to the theoretical prediction has been done for a large number of charm-anticharm-pairs, $N_{pair}=10$, mimicking a grand canonical description of the situation, as in the SHM description. On the other hand, if the number of pairs is small, the exact conservation of charm-anticharm pairs becomes relevant. In this case, the effect of canonical suppression has to be considered~\cite{Gorenstein_2001,moritz}.

\section{Conclusions and Outlook} \label{sec:conclusions}

We have developed a classical microscopic model with the goal to describe the propagation and interactions of heavy quarks in the QGP, especially the formation of bound states to account for the dissociation and regeneration of charmonia. For that purpose a relativistic Langevin equation has been used in numerical simulations to solve the Brownian motion of the heavy quarks, which move through the medium under the influence of the drag and diffusion transport coefficients. We have implemented a Coulomb-like potential between heavy particles which allows the formation of charmonium states under the condition of a negative relative energy of a charm-anticharm quark pair. The combined dynamics allows us to monitor in real time the recombination and dissociation of bound states. While these processes compete with each other, we have demonstrated that a complete thermodynamic equilibrium limit is reached eventually, in time scales dominated by the pairwise potential between the heavy particles. After tuning the strong coupling constant, the charmonium multiplicities obtained in equilibrium are in accordance with the prediction of the SHM for the low-lying charmonium states  at the critical crossover-temperature. Furthermore we have investigated the influence of the temperature, the box volume as well as the magnitude of the drag coefficient on the formation of bound states.
 %Should we also include the pz-maxwell-jüttner plot?

After having passed all the tests within box simulations and ensuring, that the correct chemical equilibrium limit is reached, and the bound state formation, dissociation and regeneration occurs in the expected manner we can now move forward and adapt our model to a situation of a realistic heavy-ion collision. In the future we will embed our model in an expanding medium, modeled by a blast-wave like elliptic fireball. Within this dynamical description the heavy quarks will initially be placed according to the Glauber model and with an initial momentum distribution given by PYTHIA. Using this setup, we can now furthermore study the elliptic flow of charm quarks and charmonia as well as the nuclear modification factor at RHIC and LHC energies. Since up to this point only regenerated charmonium states have been included in the description, it should be considered to additionally include primordial $J/\psi$ to be able to better compare to experimental data. This is especially important at RHIC energies, where regeneration processes are believed to be negligible and the dominant effect is the dissociation and suppression of initial bound states. This is work in progress. Furthermore, this description allows to be expanded to the bottom sector and the bottomonium formation, as well as $B_c$ states. Finally, since the complete description of a heavy-ion collision should also account for the hadronic medium evolution~\cite{Das:2024vac}, the whole framework could also be applied to the confined phase to describe $D$ and $\bar{D}$ mesons in a bath of light constituents, with the possibility of the $X(3872)$ formation or other molecular states at finite temperature~\cite{Montana:2022inz}. 

% \section{To be deleted}

% I take Blaizot's expression for $\gamma$ [Eq.(5.78)] and $2 \pi T D_s$ [Eq.(5.79)]. Using his parameters, $T_c=160$ MeV, $\alpha_s=0.5$, $\Lambda=4$ GeV and the expression
% \begin{equation}
% m_D^2 = \frac{16 \pi}{3} \alpha_s T^2
% \end{equation}
% I can reproduce $m_D=460$ MeV and $2 \pi TD_s =2.68$ at $T=T_c=160$ MeV. This value for $D_s$ is small, but closer to Catania model, and not as small as ours.

% Now, I changed the values of the strong coupling to $\alpha_s=0.7$, and without any other change I obtain $2 \pi TD_s =1.37$
% at $T_c=160$ MeV, confirming your result at that temperature in the plot. That means there is no problem in your calculation.

% The only inconsistency with this calculation is that if we modify $\alpha_s$, we should also modify $m_D$ using the previous formula. Then it is not $460$ MeV but $m_D=548$ MeV  ($\Lambda$ might change a bit as well...). With these values I obtain $2 \pi TD_s =1.53$, which is bit higher, but not much from what you get. 

% The conclusion is that $D_s$ is rather small in Blaizot's model, but if we---on top of that---increase $\alpha_s$ to 0.7, it gets even smaller. 

\section*{Acknowledgments}

We thank Taesoo Song for valuable discussions on the topic and for providing to us the DQPM data presented in the lower panel of Fig.~\ref{fig:drag_coefficient}.

We acknowledge support by the Deutsche Forschungsgemeinschaft (DFG) through the CRC-TR 211 “Strong- interaction matter under extreme conditions”. J.T.-R. acknowledges funding from project numbers CEX2019-000918-M (Unidad de Excelencia “María de Maeztu”), PID2020-118758GB-I00 and PID2023-147112NB-C21, financed by
the Spanish MCIN/ AEI/10.13039/501100011033/.
%\begin{acknowledgments}
%\end{acknowledgments}
	%%%%%%%%%%%%%%%%%%%%%%%%%%%%%%%%%%%%%%%%%%%%%%%%%%%%%%%%%%%%%%%%%%%%%%%%%%%%%%%
	\appendix
	
	%\section{Appendix}
	
	%%%%%%%%%%%%%%%%%%%%%%%%%%%%%%%%%%%%%%%%%%%%%%%%%%%%%%%%%%%%%%%%%%%%%%%%%%%%%%%
	
	\bibliography{references}%
	
\end{document}